\documentclass{moriond}
\usepackage{graphics}

\def\NPB{{\em Nucl. Phys.} B\ }
\def\PLB{{\em Phys. Lett.}  B\ }
\def\PRL{{\em Phys. Rev. Lett.}\ }
\def\PRD{{\em Phys. Rev.} D\ }

\def\be{\begin{equation}}
\def\ee{\end{equation}}
\def\bea{\begin{eqnarray}}
\def\eea{\end{eqnarray}}

\newcommand{\Photo}{\medskip\includegraphics[height=35mm]{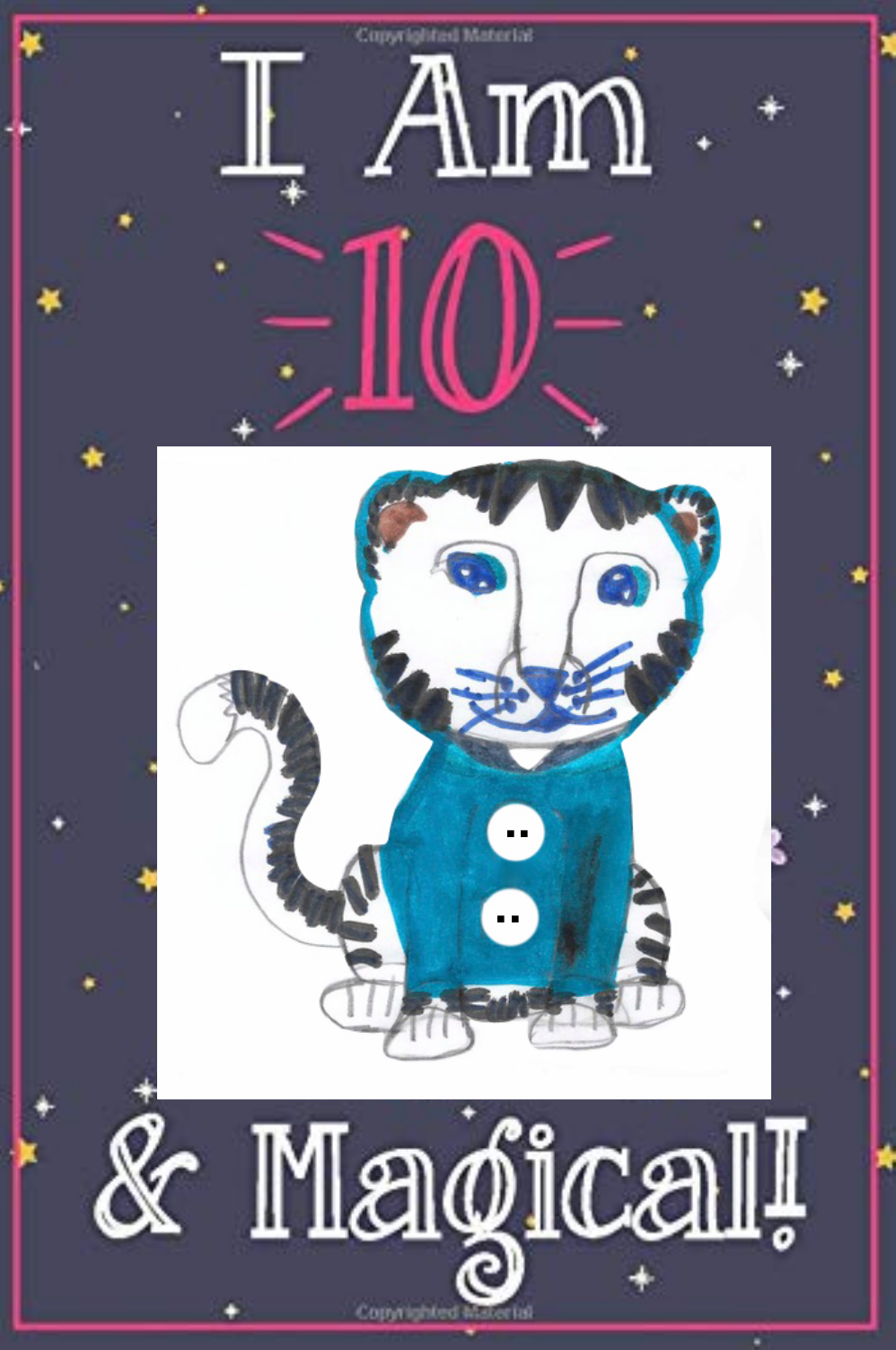}}

\begin{document}
\vspace*{4cm}
\title{HIGGS TURNS 10: THE CHILDHOOD STORY$\,$\footnote{This talk is dedicated to the great visionaries who made a theoretical breakthrough more than half-a-century ago, which took so long to appreciate and even longer to confirm experimentally: Fran\c{c}ois Englert, Carl Hagens, and Peter Higgs, and ad memoriam
Robert Brout (1928--2011), Gerald Guralnik (1936--2014), and Thomas Kibble (1932--2016).}$^,\,$\footnote{Original artwork \copyright ~Julia Landsberg, 2022.}}

\author{GREG LANDSBERG}

\address{Brown University, Department of Physics, 182 Hope St., Providence, RI 02912, USA}

\maketitle\abstracts{
In this talk, I give a historical and personal overview of the events that lead to the discovery of the Higgs
boson by the ATLAS and CMS experiments in 2012, and highlight the childhood years of the
Higgs boson studies at the Large Hadron Collider. These recollections are based on my time as the
CMS Physics Coordinator (2012--2013), and because of this they are somewhat biased toward the CMS side 
of the story, which I know first-hand.}

\section{Higgs Turns 10!}

July 4, 2022 will mark the tenth anniversary of the birth of the Higgs boson to the happy parents: the ATLAS and CMS Collaborations at the CERN Large Hadron Collider (LHC). Here are some highlights of the first ten years of the Higgs boson childhood, which I'll expand on in the subsequent sections:\\[-2pt]
\begin{minipage}{8cm}
\begin{itemize}
\item{2012:} I am born! \\[-20pt]
\item{2013:} First steps \\[-20pt]
\item{2014:} Who am I? \\[-20pt]
\item{2015:} Why am I alone?  \\[-20pt]
\item{2016:} Not afraid of dark  \\[-20pt]!
\item{2017:} Trip to the tau-land  \\[-20pt]
\item{2018:} From top to bottom  \\[-20pt]
\item{2019:} I am getting precise  \\[-20pt]
\item{2020:} Meet the 2$^{\rm nd}$ generation  \\[-20pt]
\item{2021:} How broad am I? \\[-20pt]
\item{2022:} I am charming! \\[-12pt]
\end{itemize}
\end{minipage}
\begin{minipage}{8cm}
\includegraphics[width=5cm]{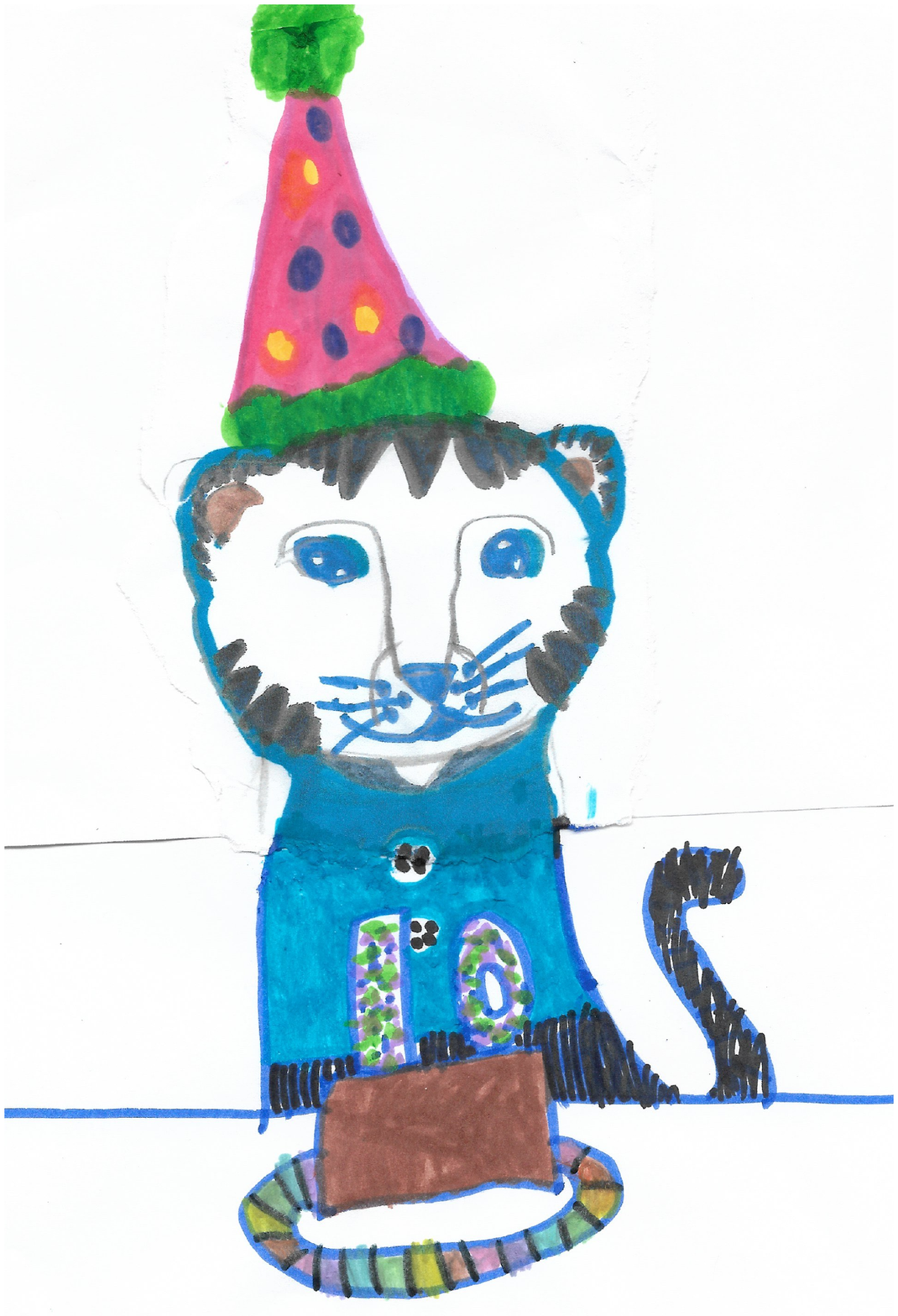}
\end{minipage}

\section{Long Road to the Higgs Boson Discovery}

The 1964 was quite a remarkable year! Many memorable events took place then: Martin Luther King, Jr. has received a peace Nobel Prize; the Beatles first arrived to America; Berkeley students stormed the university administration building, marking the Free Speech Movement; the ``Boston Strangler'', Albert DeSalvo, was arrested; and the first Ford Mustang was introduced at the World's Fair in the New York City... 

In the meantime, in physics, several remarkable discoveries took place: the quark model was established in seminal papers by Gell-Mann$\,$\cite{Gell-Mann:1964ewy} and Zweig$\,$\cite{Zweig:1964jf}; the $CP$ violation was observed by Christenson, Cronin, Fitch, and Turlay in neutral kaon decays$\,$\cite{Christenson:1964fg}, demonstrating that the time has a preferred direction; cosmic microwave background was discovered by Penzias and Wilson$\,$\cite{Penzias:1965wn}. And, hardly noticed at the time, four rather obscure papers on certain features of quantum field theory, which outlined the basics of what now known as the Brout--Englert--Higgs (BEH) mechanism, have appeared$\,$\cite{Englert:1964et,Higgs:1964ia,Higgs:1964pj,Guralnik:1964eu} as well...

In the first five years since their appearance, these four papers received only a handful of citations---sixty-four in total---while by now they have been cited altogether more than 23,000 times, solidly putting them among the most remarkable papers ever written! The real appreciation of these papers came only about a decade later, with the realization in late 1966 why the photon remains massless$\,$\cite{Kibble:1967sv}, incorporation of the BEH mechanism in the model of electroweak interactions$\,$\cite{Glashow:1961tr} by Weinberg and Salam$\,$\cite{Weinberg:1967tq,Salam:1968} in 1967, and the proof in 1971--1972 that this model, now known as the standard model (SM) of particle physics, is renormalizable, via the work of `t Hooft and Veltman$\,$\cite{tHooft:1971qjg,tHooft:1972tcz}.

The realization that the BEH mechanism must have a particle carrier, the Higgs boson, was established already in the original papers. The attempts to calculate its various properties, along with the development of the SM, spawned early phenomenological work on the Higgs boson production and decays that took place in the late 1970-ies through late 1980-ies. This work has been summarized by Gunion, Haber, Kane, and Dawson in ``The Higgs Hunter's Guide", first printed in 1989$\,$\cite{Gunion:1989we}, which instantly became a handbook for the experimentalists, as was reflected in the 2017 J.J.~Sakurai prize given to the authors. Thus, by mid 1980-ies, the experimental quest for the Higgs boson has finally started. 

Due to the lack of time I'll skip the last century searches for the Higgs boson, mainly at DORIS (DESY) and LEP (CERN). After seeing tantalizing hints of a Higgs boson at the mass of about 115~GeV, LEP was finally switched off in November 2000, to open way to the construction of the LHC. By then we learned that the Higgs boson must be rather massive: its mass must exceed 114.4~GeV at 95\% confidence level (CL)$\,$\cite{LEP}. Thus, in 2001, at the dawn of the new century, the baton was passed from Europe to the U.S., and the Fermilab Tevatron became the next place where the Higgs boson was sought. Meanwhile the construction of the LHC and the two general-purpose detectors, ATLAS and CMS, went into full swing at CERN.

The Tevatron Run II started rather slowly, and it took several years for an instantaneous luminosity to reach the design level. By mid-2005, each of the CDF and D0 experiments has accumulated an integrated luminosity of 1~fb$^{-1}$, but this amount of data was not enough to extend the LEP limits. In the meantime, the LHC construction and commissioning was completed, and the first proton beams, still at the injection energy, were circulated in the LHC on September 10, 2008. However, just nine days later, when ramping the LHC dipole magnets to their design current, a major incident happened because of a defective superconducting junction between two dipoles, which resulted in an explosion inside the LHC tunnel and a setback of the LHC startup by a whole year.

By early November 2009, while the LHC was still undergoing the repairs, the Tevatron finally published the first limits on the Higgs boson that went beyond the LEP exclusion.  Based on an integrated luminosity of between 2.1 and 5.4 fb$^{-1}$ the CDF and D0 Collaborations managed to exclude a small region of Higgs boson masses around 165~GeV$\,\,$\cite{Tevatron}. Just a few days later, on November 23, 2009, first LHC collisions at the injection energy, after the completion of the repairs, took place. The Tevatron Run II was officially extended until the end of 2011, and the race of the two machines to the ultimate goal---the discovery of the Higgs boson---has started! By March 2010, the LHC reached its initial collision energy of 7 TeV (limited by the fact that a more serious mitigation of other possibly defective junctions was required to reach the design energy). 

Meanwhile, the 2010 J.J.~Sakurai prize in theoretical physics was given to Brout, Englert, Hagen, Higgs, Guralnik, and Kibble ``for elucidation of the properties of spontaneous symmetry breaking in four-dimensional relativistic gauge theory and of the mechanism for the consistent generation of vector boson masses''. 

By summer 2011, the LHC experiments accumulated the first fb$^{-1}$ of 7 TeV data, which dramatically changed the landscape of the Higgs boson searches. Both the ATLAS and CMS Collaborations$\,$\cite{CONF-2011-112,HIG-11-022} made public their first limits on the Higgs boson mass and were able to exclude a broad range of masses from about 135 to 450 GeV. The Tevatron, in the meantime, has expanded on their exclusion to 156--177~GeV$\;$\cite{arXiv:1107.5518 }. It became clear that the spectacular LHC performance in 2011 put it way ahead of the Tevatron in the race for the Higgs boson. As a result, the request by the CDF and D0 experiments to extend the Tevatron run till 2013 was declined, and on September 30, 2011, in an emotional ``Tevatr-off" event, the Tevatron legacy of two decades has finally ended, and the machine was permanently switched off. The Higgs baton was back in the European hands.

In December 2011, ATLAS and CMS made public the last limits on the Higgs boson, based on an integrated luminosity of up to 4.9 fb$^{-1}$. The excluded range of masses was between 127 and 600 GeV$\;$\cite{HIG-11-032,ATLAS:2012ae} . More importantly both collaborations saw some excesses in the mass range of 124--126 GeV, but it was not clear if they are compatible between different channels analyzed and between the two experiments. From then on, ATLAS and CMS focused on the discovery of the Higgs boson in the low-mass range, around 115--130 GeV.

In CMS, we decided to bet on the fact that the machine energy will be increased to 8 TeV for 2012 running, and started producing large Monte Carlo event samples for this energy as early as in November 2011, which were necessary for proper optimization of the Higgs boson search and detailed understanding of the backgrounds. This bet played out very well and actually allowed CMS to show all five main decay channels of the Higgs boson: $H \to ZZ \to 4\ell$, $H \to \gamma\gamma$, $H \to WW$, $H \to \tau\tau$, and $H \to b\bar b$ at the time of the discovery. The decision to increase the machine energy was not easy. The LHC machine physicists wanted to err on the side of caution and run at the same 7~TeV energy as before; the ATLAS and CMS experiments were pushing for a safe increase in energy, as this would minimize the amount of time needed for the discovery. At the end, the experiments prevailed, and at the beginning of February 2012, at the annual Chamonix LHC workshop, it was decided to operate the machine at 8 TeV. The rest is now history!

The LHC startup in 2012 was very smooth and already early in the run it became clear that the machine will double the amount of the delivered integrated luminosity  by early summer. Half of this integrated luminosity was to be delivered at higher energy, with the dominant Higgs boson gluon fusion cross section 30\% higher than at 7 TeV. It was therefore decided to plan for the next major update on the Higgs boson searches for the ICHEP conference, scheduled to start on July 4, 2012 in Melbourne. It was anticipated that if the $\sim$125 GeV excess seen by both ATLAS and CMS experiments is real, the new data would suffice to establish the existence of the Higgs boson at the discovery level, i.e., five standard deviations in each experiment.
 
The analysis of the complexity of the Higgs boson search was performed in a ``blind" way, i.e., the data in the mass region of 110--140 GeV was kept hidden in order not to subconsciously bias the results. The analysis was fully optimized, and the backgrounds and all the systematic uncertainties were estimated with the data in the region of interest still being blinded. By mid-June 2012, CMS was ready to finally unblind the analyses in all five main decay channels. 

\noindent
 \begin{minipage}{12cm}
\hspace{16pt}
The unblinding of the data was done as a formal event on Friday, June 15, 2012, in the Filtration Plant auditorium at CERN and via video. 
The entire collaboration was invited to join the event. About 250 people were sitting and standing in the room for about 2 hours it took to show the results in all five channels, on a hot summer day with the temperature in the room reaching 28$^\circ$ C, but nobody minded! More than 500 more people were connected to the event via video. 
\end{minipage}
\begin{minipage}{4cm}
\includegraphics[width=3.5cm]{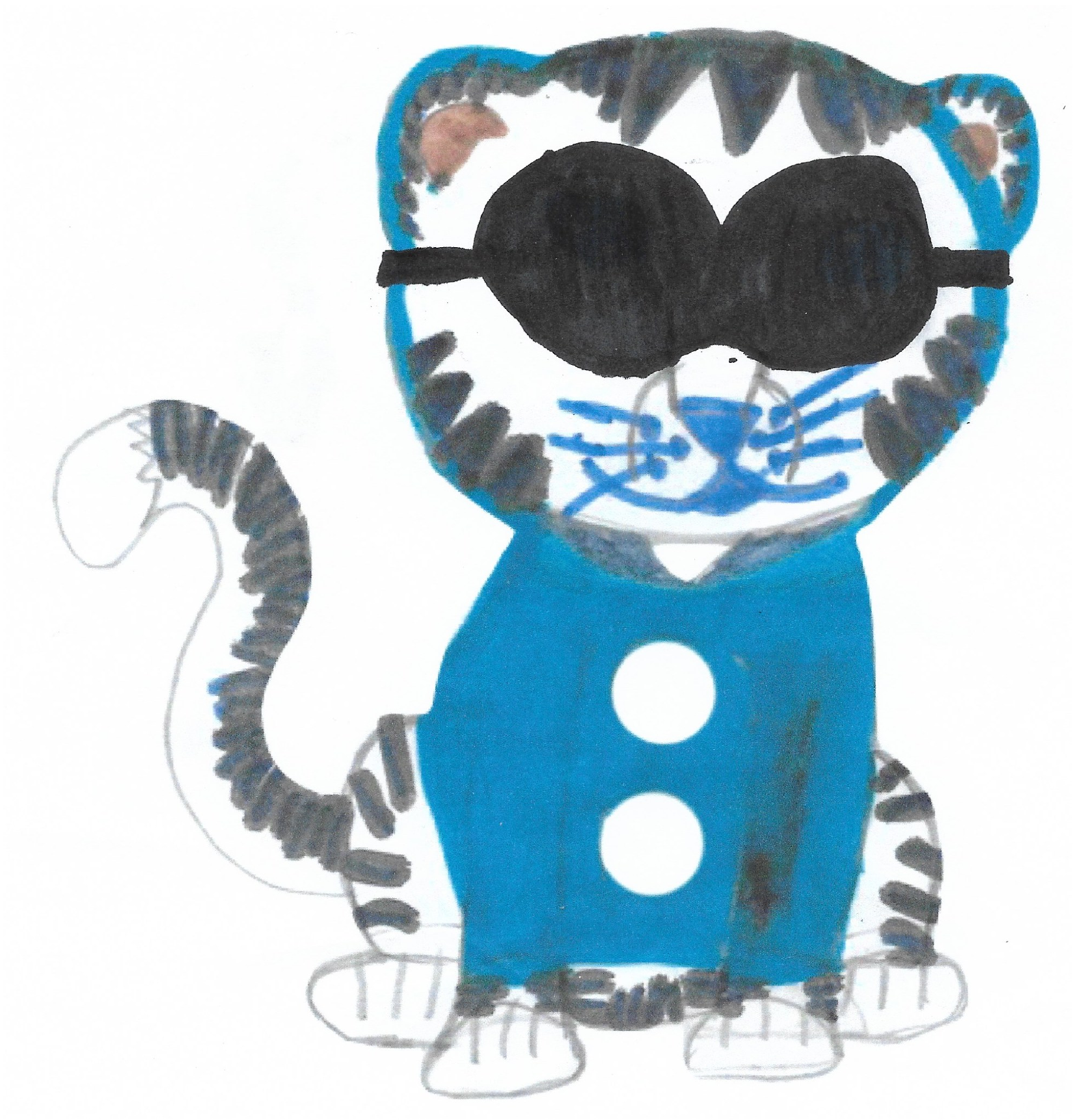}
\end{minipage}

The unblinding was done based on about 80\% of the full data set. The most sensitive $\gamma\gamma$ and $4\ell$ channels were unblinded first, and to the cheers of the audience we have observed very clear peaks at 125 GeV in both of these channels with the significances around four standard deviations each! We knew then that we had the Higgs boson; the rest of the world will learn about it three weeks later...

The next day, on Saturday June 16, I decided to throw an informal discovery party at our place in Challex, France. Some 30 people attended, but given that I had several ATLAS colleagues leaving nearby, we ``masqueraded" this party as a birthday party, complete with a cake proudly displaying ``125" with candles. Still, it was probably a big surprise for the trash collectors the following Tuesday to find a recycling bin filled with about 30 empty champagne bottles next to my house. The Higgs was born!

\begin{center}
\noindent\includegraphics[width=0.85\textwidth]{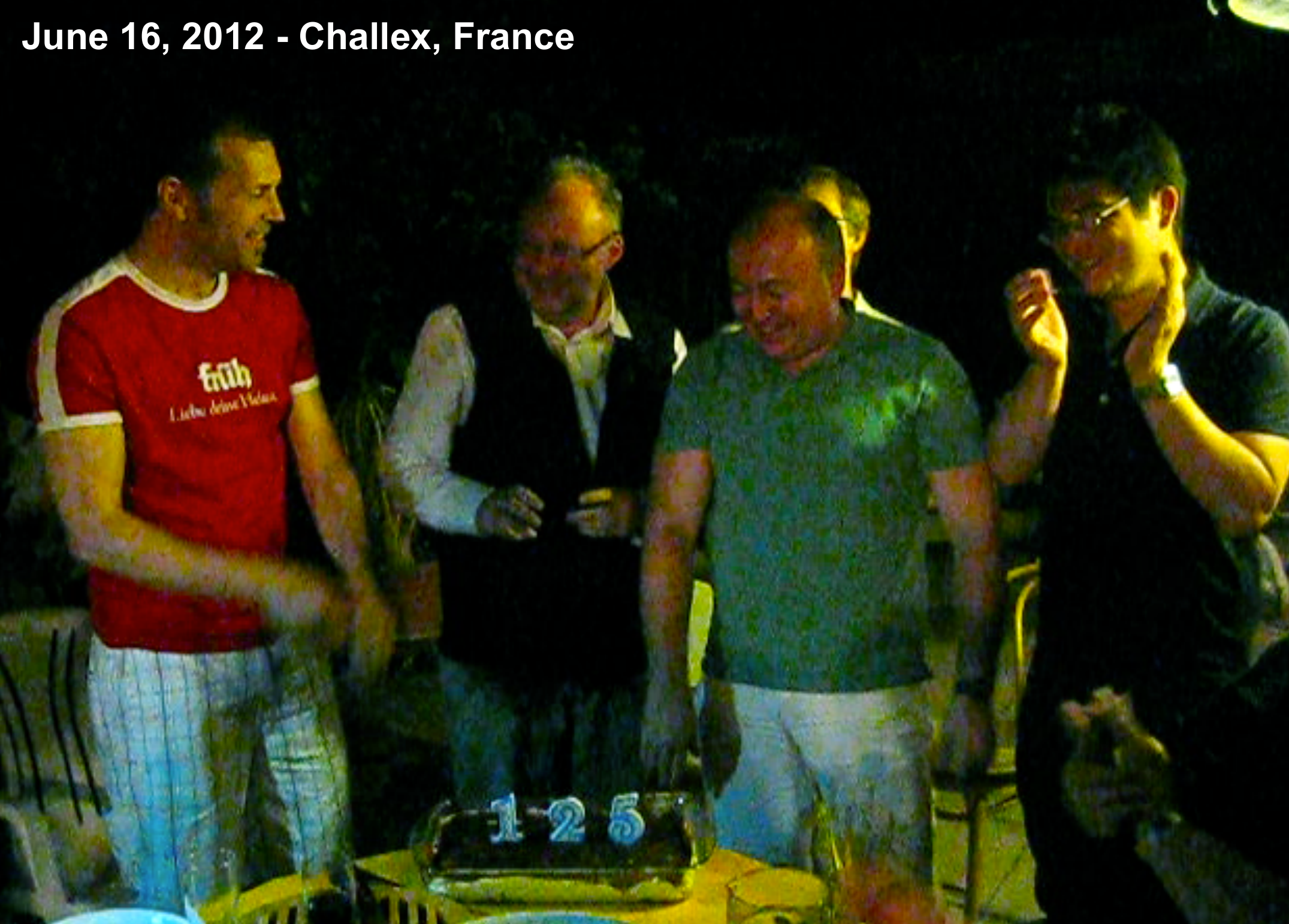}
\end{center}

The following three weeks were a mad rush to finalize everything for the discovery announcement. The rest of the data were added, with the analysis selections completely frozen at the time of the unblinding, and the five channels were combined together to yield the overall observed (expected) significance of 5.0 (5.8) standard deviations. The discovery presentation was scheduled for July 4th, in the CERN Main Auditorium, simultaneously broadcasted to Melbourne. The ATLAS and CMS spokespeople, Fabiola Gianotti and Joe Incandela, presented the results to the world. Each experiment achieved 5 standard deviations---the gold standard for the discovery---independently. At the end of the presentation, Rolf-Dieter Heuer, the CERN Director General, said: ``I think we have it!"  The Higgs was now born officially! Given the Independence Day holiday in the U.S., the fireworks came for free...

The two discovery papers were submitted to the Physics Letters B journal on July 31, 2012 and appeared online on August 18$\;$\cite{ATLAS-Higgs,CMS-Higgs}.

\section{Higgs at Moriond}

\begin{minipage}{9cm}
More data were delivered by the LHC in 2012, and the discovery sample was more than doubled. The next big update on the Higgs boson was to be announced at the Moriond series of conferences, in March 2013. However, in CMS, we saw that the significance of the Higgs boson signal in the very sensitive diphoton channel did drop despite significantly more data analyzed. This was very puzzling, as nothing has changed in the analysis, and the significance in the most sensitive four-lepton channel kept increasing.
\end{minipage}
\begin{minipage}{8cm}
\includegraphics[width=7cm]{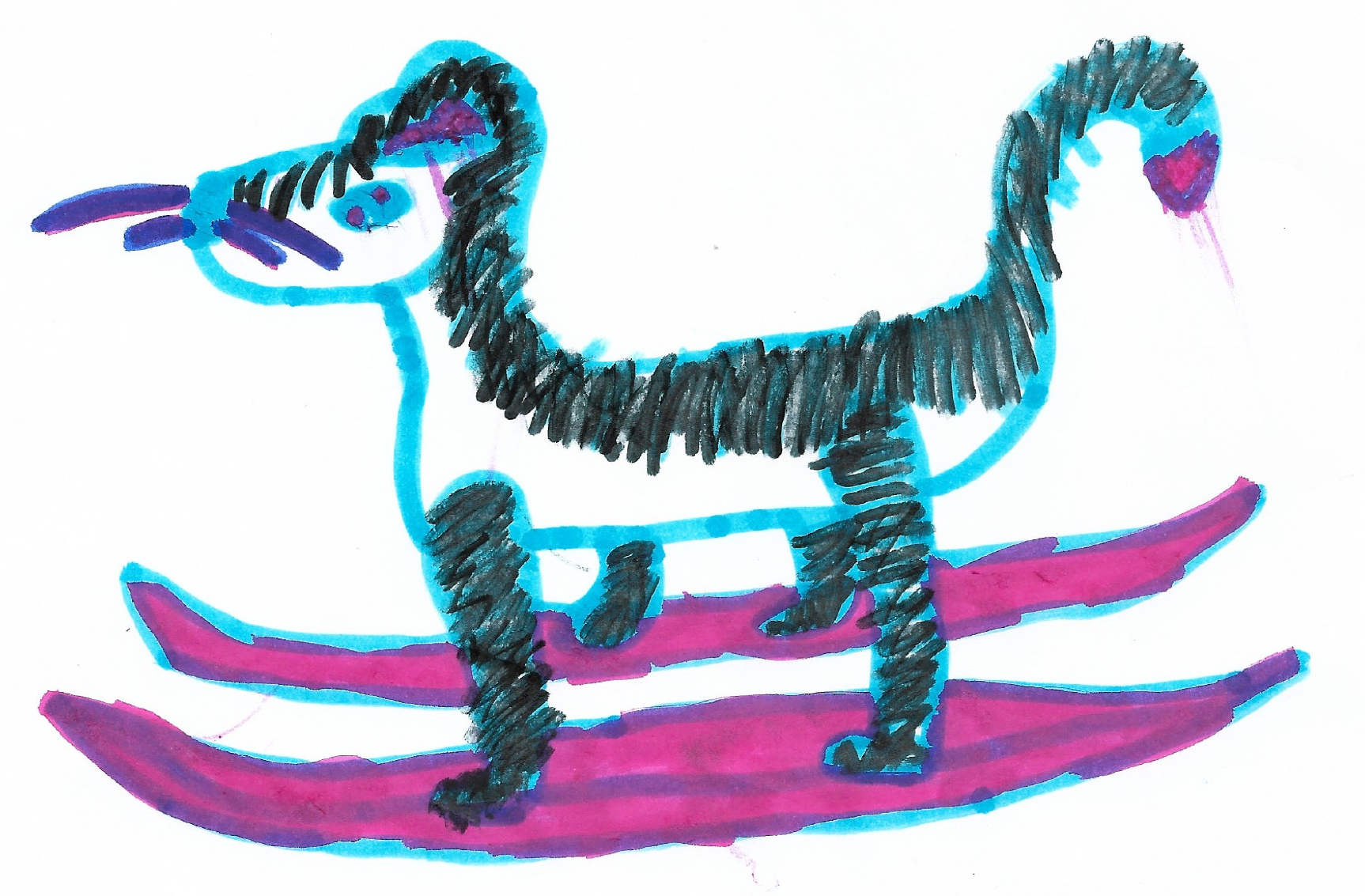}
\end{minipage}

\smallskip
 It took us several months to decisively prove that there were no mistakes in the analysis and that the drop in the significance was a result of a statistical fluctuation. The calibration of the electromagnetic calorimeter, which is crucial for the high-resolution diphoton channel, has been improved, and that was the main change that resulted in the statistical fluctuation. When dealing with the case of a rather large signal on top of an overwhelming background, the efficiency of the optimal selection is fairly low. Even small changes in the kinematic parameters of the events, such as the change in the photon energy due to the new calibration, could result in completely different signal candidates to be selected. Indeed, when the new calibration was applied to the discovery sample, we noticed that the overlap between the signal candidates in the discovery analysis and in the new analysis was not that large; in fact we were dealing with largely statistically independent event candidate samples in the two analyses. In such a situation, large statistical fluctuation are possible. Using a special statistical technique, known as the jack-knife resampling$\,$\cite{jack-knife}, we were able to show that the two results were in fact consistent with one the other within 2 standard deviations and approved the new diphoton result for presentation at the Moriond QCD conference. We practiced the talk after the dinner on the night before the presentation, well beyond midnight, since we have expected a lot of questions about this new result. The observed (expected) significance of the Higgs boson signal was 3.2 (4.2) standard deviations, a big change from the 4.1 (2.8) standard deviations at the time of the discovery.  Statistics could play funny tricks! As a result of the late-night practice talk, some of us, including myself, were half-asleep during the morning session where the new result was shown for the first time, which was captured by the sharp eye of the Moriond photographer!

That Moriond conference was quite memorable, as the host of new results released by the ATLAS and CMS experiments demonstrated with an increased precision that the new particle was indeed consistent with the SM Higgs boson. The CERN press release stated: ``New results indicate that a new particle is a Higgs boson"! Since then we dropped the term ``Higgs-like particle", which was originally used to denote the new boson, and started calling it ``a Higgs boson" and, a bit later, ``the Higgs boson". Incidentally, the same week the new pope---Pope Francis---was confirmed in Vatican. The Simon Fraser University (a member of the ATLAS Collaboration) newsletter connected these two events on March 14, 2013 under the title ``New boson and new pope confirmed".

\section{Dazzling Precision}

We have learned a lot about the Higgs boson in the past ten years, thanks to the spectacular performance of the LHC machine, which delivered 15 times the discovery sample, most of it at much higher energy of 13 TeV. This rapid progress would not have been possible without significant theoretical work on the Higgs physics, captured in four Yellow Reports of the LHC Higgs Cross Section Working Group, as well as in more recent papers. Many new results based on this large data set have been shown at this conference.

At the time of the discovery the Higgs boson production cross section in various decay channels in each experiment was measured with the precision of about 25\%. The precision has now reached 6\%$\;$\cite{ATLAS-CONF-2021-053}. By 2015, just a couple of years after the discovery, we have firmly established that the Higgs boson has quantum numbers of the vacuum $J^{PC} = 0^{++}$ and ruled out pure pseudoscalar or tensor hypotheses at $>$99.9\% CL$\,$\cite{ATLAS-CP,CMS-CP}. (The fact that Higgs boson can't be a spin-1 particle is evident from its decay to a pair of photons, which is not possible for a spin-1 particle, according to the Landau--Yang theorem.)

\noindent
\begin{minipage}{3.6cm}
\includegraphics[width=3.6cm]{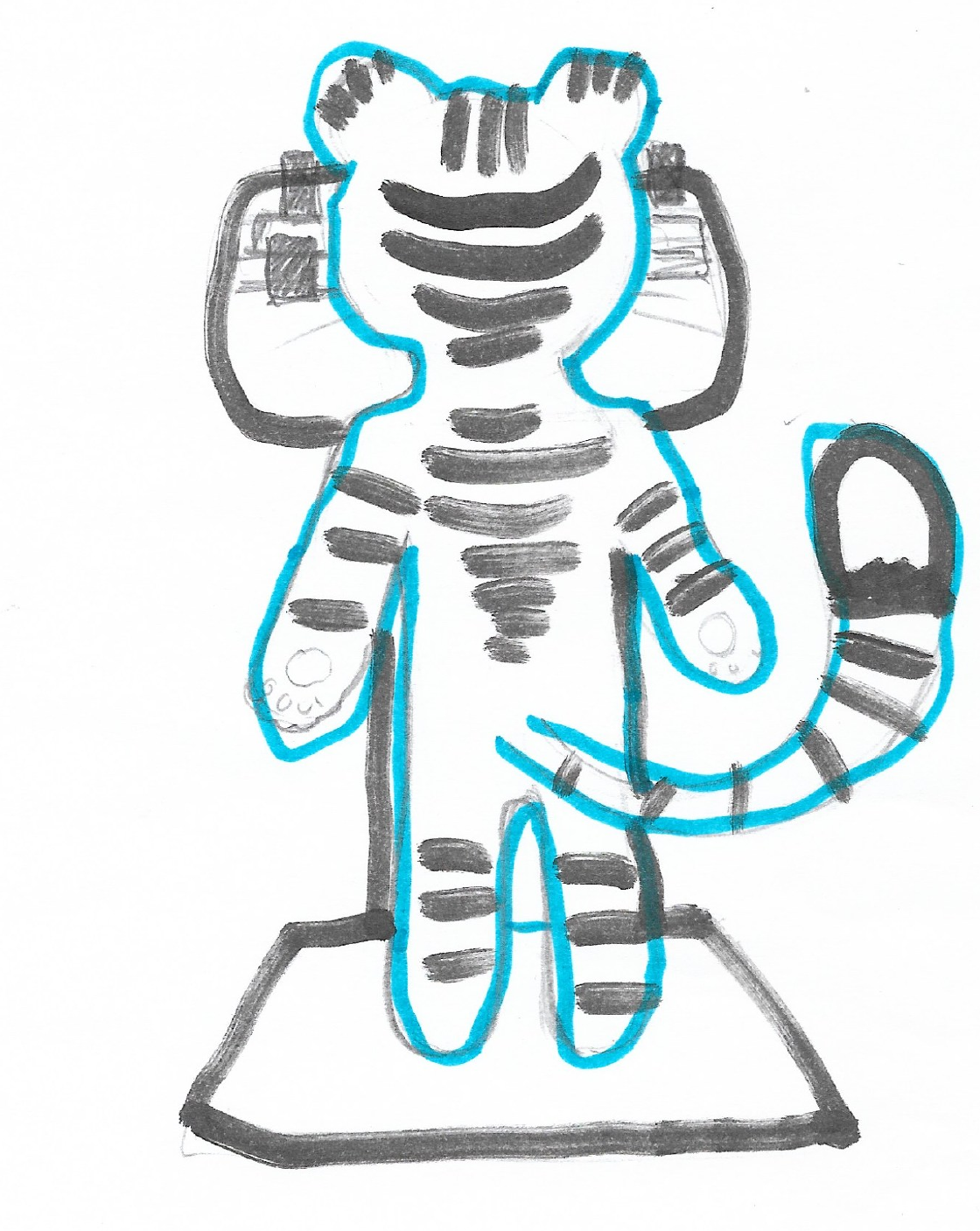}
\end{minipage}
\begin{minipage}{8.9cm}
\hspace{16pt}
After a slight initial tension between the two experiments on the value of the Higgs boson mass, it is now measured to a per mil precision. The most precise measurement to date, coming from the CMS experiment, is $125.38 \pm 0.14$ GeV$\;$\cite{CMS-Higgs-Mass}. Very recently, a two-sided 95\% CL  interval on the Higgs boson width and the first evidence for the off-shell Higgs boson production were reported, with the measured width $\Gamma_H = 3.2 ^{+2.4}_{-1.7}$ GeV$\;$\cite{CMS-Higgs-Width} (fully consistent with the width of 4.1 MeV predicted in the SM).
\end{minipage}
\begin{minipage}{3.6cm}
\includegraphics[width=3.6cm]{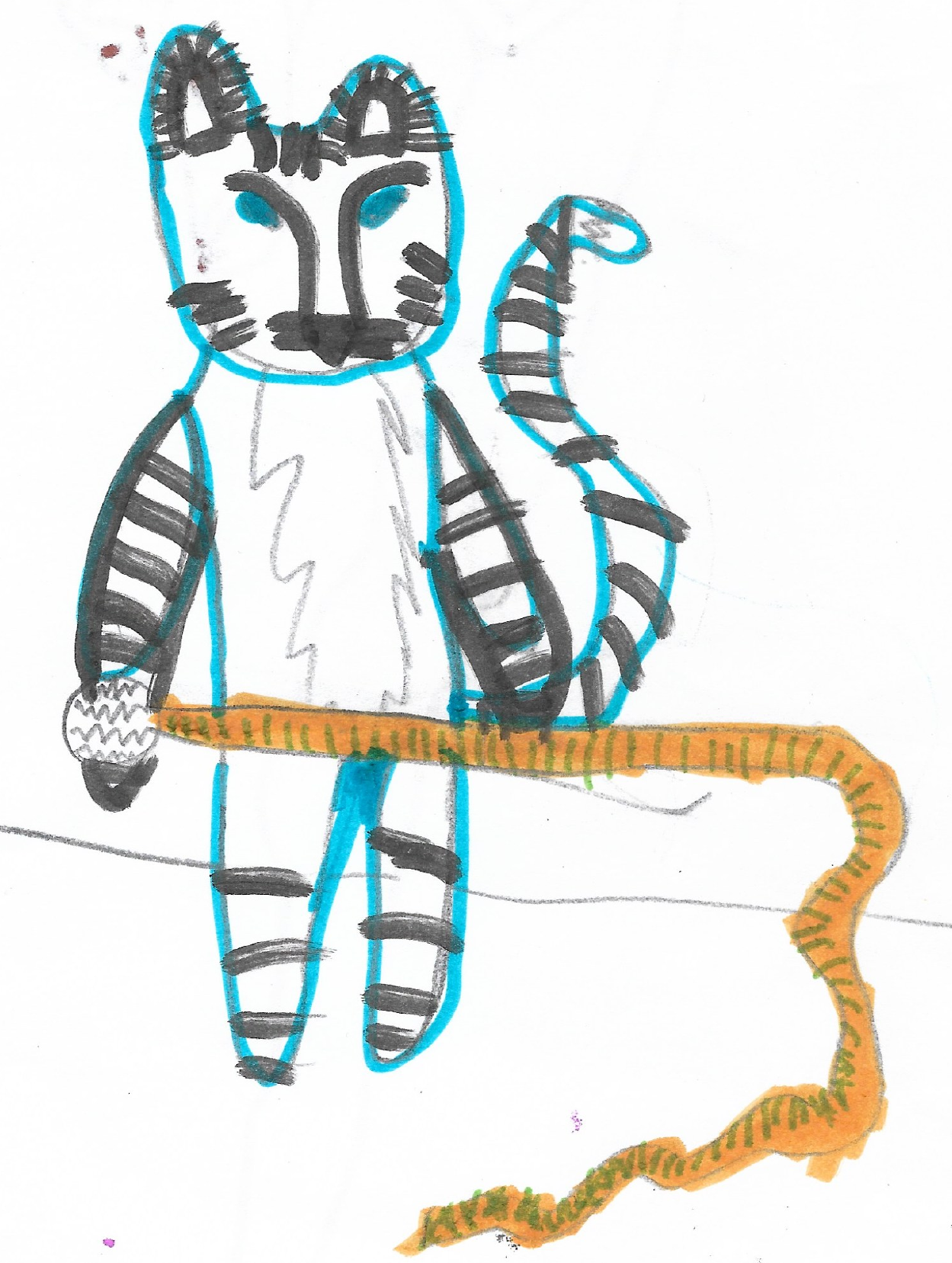}
\end{minipage}

By 2018, the coupling of the Higgs boson to $\tau$ leptons has been firmly established by both ATLAS$\,$\cite{ATLAS-tau} and CMS$\,$\cite{CMS-tau}, with the significance around 6 standard deviations in each experiment, thus establishing the coupling to the third-generation quarks (indirectly, via the gluon fusion production through a top quark loop) and leptons. Later the same year, the direct coupling to top quarks via the $t\bar t H$ production was established as well$\,$\cite{ATLAS-top,CMS-top}. And finally, also in 2018, the coupling of the Higgs boson to bottom quarks was finally measured by both collaborations, primarily via associated production with vector bosons, with an overall significance of about 5.5 standard deviations by each experiment$\,$\cite{ATLAS-b,CMS-b}. Ironically, this decay with the largest branching fraction among all other Higgs boson decays took six years and tour-de-force analyses to be finally observed, since it suffers from very large backgrounds. The 2018 became the year of the third generation!

In 2019, first evidence for Higgs boson decay to a second-generation fermion (the muon) was seen by CMS at the 3 standard deviations level$\;$\cite{CMS-mu}; ATLAS also saw an excess of dimuon events at the Higgs boson mass corresponding to 2 standard deviations$\,$\cite{ATLAS-mu}. Finally, in 2022, both ATLAS and CMS presented impressive limits on the coupling of 
Higgs bosons to charm quarks, which, while still being far from the SM expectations, showed that it may be possible to establish this decay mode at the High-Luminosity LHC.

In parallel with this quest for various SM decay channels of the Higgs boson, both ATLAS and CMS looked for other, exotic decay channels, which are either prohibited or very suppressed in the SM. Among these decays are the lepton flavor violating decays, such as $H \to \tau\mu$, as well as invisible decays of the Higgs boson, which allow to probe potential connection between the Higgs boson and dark matter. The most recent result from ATLAS$\,$\cite{Higgs-DM} has established a fairly stringent upper limit on the branching fraction of invisible Higgs boson decays at 14.5\% at 95\% CL (with the 10.3\% expected limit).

Four dedicated experimental talks at this conference, along with three theoretical ones, showcased some of these results, as well as discussed future measurements, including the pair production of Higgs bosons.

\section{Conclusions}

Higgs boson had a fun childhood and is now finishing the ``primary school". The ATLAS and CMS experiments have been caring and watchful parents for all these childhood years, and will follow the ``middle school" years with the large Run 3 and beyond data samples. With the Higgs boson couplings to third-generation fermions well established, our sights are shifting on the second generation and on exotic decays, as well as on precision differential measurements and on the Higgs boson pair production. Happy birthday, the Higgs boson!

\section*{Acknowledgments}

I'm grateful to the organizers for inviting me to give this talk, which jogged a lot of fond memories. This work is partially supported by the U.S. Department of Energy grant no. DE-SC0010010. Special thanks to my daughter Julia, who prepared a nice set of drawings for this talk with an artist's rendition of its main character, some of which made it to these proceedings.

\end{document}